%% file: template_nika2.tex
\documentclass{webofc}
\usepackage[varg]{txfonts}   
\begin{document}
\title{Crab nebula at 260 GHz with the NIKA2 polarimeter}
%
%
\subtitle{Implications for the polarization angle calibration of future CMB experiments}

\author{\firstname{A.}~\lastname{Ritacco}\inst{\ref{IAS},\ref{ENS}}\fnsep\thanks{\email{alessia.ritacco@phys.ens.fr}}
  \and \firstname{R.}~\lastname{Adam} \inst{\ref{LLR}}
  \and  \firstname{P.}~\lastname{Ade} \inst{\ref{Cardiff}}
  \and  \firstname{H.}~\lastname{Ajeddig} \inst{\ref{CEA}}
  \and  \firstname{P.}~\lastname{Andr\'e} \inst{\ref{CEA}}
  \and \firstname{E.}~\lastname{Artis} \inst{\ref{LPSC}}
  \and \firstname{J.}~\lastname{Aumont} \inst{\ref{IRAP}}
  \and  \firstname{H.}~\lastname{Aussel} \inst{\ref{CEA}}
  \and  \firstname{A.}~\lastname{Beelen} \inst{\ref{IAS}}
  \and  \firstname{A.}~\lastname{Beno\^it} \inst{\ref{Neel}}
  \and  \firstname{S.}~\lastname{Berta} \inst{\ref{IRAMF}}
  \and  \firstname{L.}~\lastname{Bing} \inst{\ref{LAM}}
  \and  \firstname{O.}~\lastname{Bourrion} \inst{\ref{LPSC}}
  \and  \firstname{M.}~\lastname{Calvo} \inst{\ref{Neel}}
  \and  \firstname{A.}~\lastname{Catalano} \inst{\ref{LPSC}}
  \and  \firstname{M.}~\lastname{De~Petris} \inst{\ref{Roma}}
  \and  \firstname{F.-X.}~\lastname{D\'esert} \inst{\ref{IPAG}}
  \and  \firstname{S.}~\lastname{Doyle} \inst{\ref{Cardiff}}
  \and  \firstname{E.~F.~C.}~\lastname{Driessen} \inst{\ref{IRAMF}}
  \and  \firstname{A.}~\lastname{Gomez} \inst{\ref{CAB}}
  \and  \firstname{J.}~\lastname{Goupy} \inst{\ref{Neel}}
  \and  \firstname{F.}~\lastname{K\'eruzor\'e} \inst{\ref{LPSC}}
  \and  \firstname{C.}~\lastname{Kramer} \inst{\ref{IRAME}}
  \and  \firstname{B.}~\lastname{Ladjelate} \inst{\ref{IRAME}}
  \and  \firstname{G.}~\lastname{Lagache} \inst{\ref{LAM}}
  \and  \firstname{S.}~\lastname{Leclercq} \inst{\ref{IRAMF}}
  \and  \firstname{J.-F.}~\lastname{Lestrade} \inst{\ref{LERMA}}
  \and  \firstname{J.-F.}~\lastname{Mac\'ias-P\'erez} \inst{\ref{LPSC}}
  \and  \firstname{A.}~\lastname{Maury} \inst{\ref{CEA}}
  \and  \firstname{P.}~\lastname{Mauskopf} \inst{\ref{Cardiff},\ref{Arizona}}
  \and  \firstname{F.}~\lastname{Mayet} \inst{\ref{LPSC}}
  \and  \firstname{A.}~\lastname{Monfardini} \inst{\ref{Neel}}
  \and  \firstname{M.}~\lastname{Mu\~noz-Echeverr\'ia} \inst{\ref{LPSC}}
  \and  \firstname{L.}~\lastname{Perotto} \inst{\ref{LPSC}}
  \and  \firstname{G.}~\lastname{Pisano} \inst{\ref{Cardiff}}
  \and  \firstname{N.}~\lastname{Ponthieu} \inst{\ref{IPAG}}
  \and  \firstname{V.}~\lastname{Rev\'eret} \inst{\ref{CEA}}
  \and  \firstname{A.~J.}~\lastname{Rigby} \inst{\ref{Cardiff}}
  \and  \firstname{C.}~\lastname{Romero} \inst{\ref{Pennsylvanie}}
  \and  \firstname{H.}~\lastname{Roussel} \inst{\ref{IAP}}
  \and  \firstname{F.}~\lastname{Ruppin} \inst{\ref{MIT}}
  \and  \firstname{K.}~\lastname{Schuster} \inst{\ref{IRAMF}}
  \and  \firstname{S.}~\lastname{Shu} \inst{\ref{Caltech}}
  \and  \firstname{A.}~\lastname{Sievers} \inst{\ref{IRAME}}
  \and  \firstname{C.}~\lastname{Tucker} \inst{\ref{Cardiff}}
  \and  \firstname{R.}~\lastname{Zylka} \inst{\ref{IRAMF}}
}

\institute{
  Institut d'Astrophysique Spatiale (IAS), CNRS, Universit\'e Paris Sud, Orsay, France
  \label{IAS}
  \and
  Laboratoire de Physique de l’\'Ecole Normale Sup\'erieure, ENS, PSL Research University, CNRS, Sorbonne Universit\'e, Universit\'e de Paris, 75005 Paris, France 
  \label{ENS}
  \and
  LLR (Laboratoire Leprince-Ringuet), CNRS, École Polytechnique, Institut Polytechnique de Paris, Palaiseau, France
  \label{LLR}
  \and
  School of Physics and Astronomy, Cardiff University, Queen’s Buildings, The Parade, Cardiff, CF24 3AA, UK 
  \label{Cardiff}
  \and
  AIM, CEA, CNRS, Universit\'e Paris-Saclay, Universit\'e Paris Diderot, Sorbonne Paris Cit\'e, 91191 Gif-sur-Yvette, France
  \label{CEA}
  \and
  Univ. Grenoble Alpes, CNRS, Grenoble INP, LPSC-IN2P3, 53, avenue des Martyrs, 38000 Grenoble, France
  \label{LPSC}
  \and
  CNRS, IRAP, 9 Av. colonel Roche, BP 44346, 31028 Toulouse cedex 4, France
  \label{IRAP}
  \and
  Institut N\'eel, CNRS, Universit\'e Grenoble Alpes, France
  \label{Neel}
  \and
  Institut de RadioAstronomie Millim\'etrique (IRAM), Grenoble, France
  \label{IRAMF}
  \and
  Aix Marseille Univ, CNRS, CNES, LAM (Laboratoire d'Astrophysique de Marseille), Marseille, France
  \label{LAM}
  \and 
  Dipartimento di Fisica, Sapienza Universit\`a di Roma, Piazzale Aldo Moro 5, I-00185 Roma, Italy
  \label{Roma}
  \and
  Univ. Grenoble Alpes, CNRS, IPAG, 38000 Grenoble, France 
  \label{IPAG}
  \and
  Centro de Astrobiolog\'ia (CSIC-INTA), Torrej\'on de Ardoz, 28850 Madrid, Spain
  \label{CAB}
  \and  
  Instituto de Radioastronom\'ia Milim\'etrica (IRAM), Granada, Spain
  \label{IRAME}
  \and 
  LERMA, Observatoire de Paris, PSL Research University, CNRS, Sorbonne Universit\'e, UPMC, 75014 Paris, France  
  \label{LERMA}
  \and
  School of Earth and Space Exploration and Department of Physics, Arizona State University, Tempe, AZ 85287, USA
  \label{Arizona}
  \and 
  Department of Physics and Astronomy, University of Pennsylvania, 209 South 33rd Street, Philadelphia, PA, 19104, USA
  \label{Pennsylvanie}
  \and 
  Institut d'Astrophysique de Paris, CNRS (UMR7095), 98 bis boulevard Arago, 75014 Paris, France
  \label{IAP}
  \and 
  Kavli Institute for Astrophysics and Space Research, Massachusetts Institute of Technology, Cambridge, MA 02139, USA
  \label{MIT}
  \and
  Caltech, Pasadena, CA 91125, USA
  \label{Caltech}
}

\abstract{%
  The quest for primordial gravitational waves enclosed in the Cosmic Microwave Background (CMB) polarization B-modes signal motivates the development of a new generation of high sensitive experiments (e.g. CMB-S4, LiteBIRD) that would allow them to detect its imprint.Neverthless, this will be only possible by ensuring a high control of the instrumental systematic effects and an accurate absolute calibration of the polarization angle.
  The Crab nebula is known to be a polarization calibrator on the sky for CMB experiments, already used for the Planck satellite it exhibits a high polarized signal at microwave wavelengths. In this work we present Crab polarization observations obtained at the central frequency of 260 GHz with the NIKA2 instrument and discuss the accuracy needed on such a measurement to improve the constraints on the absolute angle calibration for CMB experiments.}
\maketitle
\section{Introduction}
\label{intro}
Cosmic Microwave Background (CMB) observations have allowed us to establish the current standard model for cosmology, a.k.a the $\Lambda$CDM model. The inflation theory \cite{guth1981, LINDE1982} has been introduced to explain an observational evidence that the Universe appears to be almost exactly Euclidean and uniform in all directions. This theory postulates that the Universe exponentially expanded in a tiny fraction of a second in the early times and predicts the existence of a relic background of primordial gravitational waves, which produces a distinct, curl-like signature in the polarization of the CMB, a.k.a CMB B-modes. The search for primordial gravitational waves B-modes, which are a direct probe of inflation, is one of the main ambitious goals of the modern cosmology and it motivates the development of several facilities from ground (e.g. BICEP3+Keck, CLASS, SO, SPT3G, S4, AdvACTPOL, QUIJOTE), balloon-borne to space experiments (PIPER, PICO, LiteBIRD).
These experiments share the goal of detecting the tensor-to-scalar ratio {\it r}, which gives the relative amplitude of the primordial tensor (B-modes) and scalar perturbations (E-modes) of the CMB. This parameter gives the energy scale of inflation and is expected at a range of amplitudes between 10$^{-2}$ to 10$^{-4}$ \citep{litebird2021, Xu_2021}.

In order to reach the sensitivity required for an unambiguous detection of the primordial B-modes we need to: increase the signal-to-noise ratio by scaling up the number of detectors; use optical systems as half-wave-plates to modulate the sky polarization signal and filter out low frequency noise; reduce residual systematic effects \citep{ritacco2017, odea}. Furthermore the determination of the polarization angle absolute calibration is crucial to ensure an unbiased detection of the polarization orientation of the CMB.

A recent study \citep{aumont2020,crabnika} addressed this point by using the brightest polarized astrophysical object in the microwave sky, the Crab nebula (or Tau A). They point out that the polarization angle is constant and stable in a frequency range of 23-353 GHz consisting with a mean angle of -88.26$^{\circ}$ $\pm$ 0.27$^{\circ}$ (Galactic coordinates), which would allow us to calibrate experiments with an accuracy enabling the measurement of {\it r} $\sim$ 0.01, not yet enough at the level required.

In the following we present the preliminary results from observations of the Crab nebula obtained with NIKA2 during the commissioning campaign of November 2020 and we discuss how improving the precision on these measurements could allow us to reach the sensitivity requirement for the detection of the primordial CMB B-modes. 

\section{NIKA2 observations}
\label{sec-1}
The NIKA2 instrument \cite{NIKA2-general} performes high angular resolution simultaneous observations at 1.15 and 2.05 mm, from the IRAM 30m telescope, being the 1.15 mm (260 GHz) channel also able to detect the linear polarization. In total intensity it is operating for open time observations since 2017 \cite{NIKA2-performance}, when it also started the commissioning phase of its polarization facilities. The NIKA2 polarimeter consists of a continuously rotating Half-Wave-Plate (HWP) \cite{ritacco2017} placed at ambient temperature in front of the NIKA2 entrance. The detection of the linear polarization expressed through the Stokes parameters Q and U is possible thanks to two detector arrays placed at ~100 mK in the 260 GHz channel inside the cryostat \cite{NIKA2-instrument}. For more details on the polarization system see \cite{ritacco2017}.

\subsection{Crab nebula observations}
\label{sub sec-2}
The Crab nebula has an extension of 5x7 arcminutes in the sky and it exhibits a high polarization, of the order of $\sim$ 20\% of the total intensity at the peak emission position of the nebula. \cite{crabnika} presented a compendium of all avalaible observations in a large range of frequencies (23-353 GHz) and demonstrated that the synchrotron emission is responsible for both the total intensity and polarization observed. Moreover they found a strong case for stability of the polarization angle, which places the Crab nebula as a good candidate for a polarization angle calibrator for CMB experiments. However at higher frequencies (above 200 GHz) the sample is smaller contributing to increase the uncertainty on the angle stability.

Observations performed with the NIKA2 polarimeter would increase the precision on the polarization angle estimation. Observations on the the Crab nebula have been performed with the NIKA2 instrument during three commissioning campaigns, on December 2018, February and November 2020. The polarization Stokes Q and U maps obtained from 70 minutes of observation performed during the campaign of November 2020 are shown in figure~1. 

The left panel of figure~2 shows the Crab polarization angles estimated during the three campaigns. The averaged polarization angle (solid black line) is computed as median of the polarization angles, estimated as $\psi_{Gal} = 0.5 \arctan{U/Q}$ (where the Stokes Q and U have been computed by integrating the emission flux over a 7$^\prime$ diameter where SNR > 3 in Stokes Q and U maps) of the three data sets and corresponds to -87.195 ± 0.806$^{\circ}$, expressed in galactic coordinates. Notice that the angles have been corrected for a -5.6 $\pm$ 1 $^{\circ}$ offset in absolute angle calibration on the sky as measured in OMC1 and 3C286 targets, being 1$^{\circ}$ the uncertainty accounting for the stability of this offset over almost two years of observations. Taking only the November 2020 data, which are the most reliable across the two years we obtain the median value of $\psi$ = -87.23 $\pm$ 0.29 deg. On the right panel of the figure we compare this result to other experiments \cite{crabnika}.
Making the assumption that in the future we will be able to constrain the absolute offset angle and reduce the 1$^{\circ}$ uncertainty to negligible values, we use the November data to calculate the total weighted polarization angle (including the results of all experiments). In this case we would obtain $\psi$ = -87.55 $\pm$ 0.22 $^{\circ}$, which would strongly reduce the uncertainty on the polarization angle reconstruction at higher frequencies.


\begin{figure*}[h]
\begin{center}
  \includegraphics[scale=0.1]{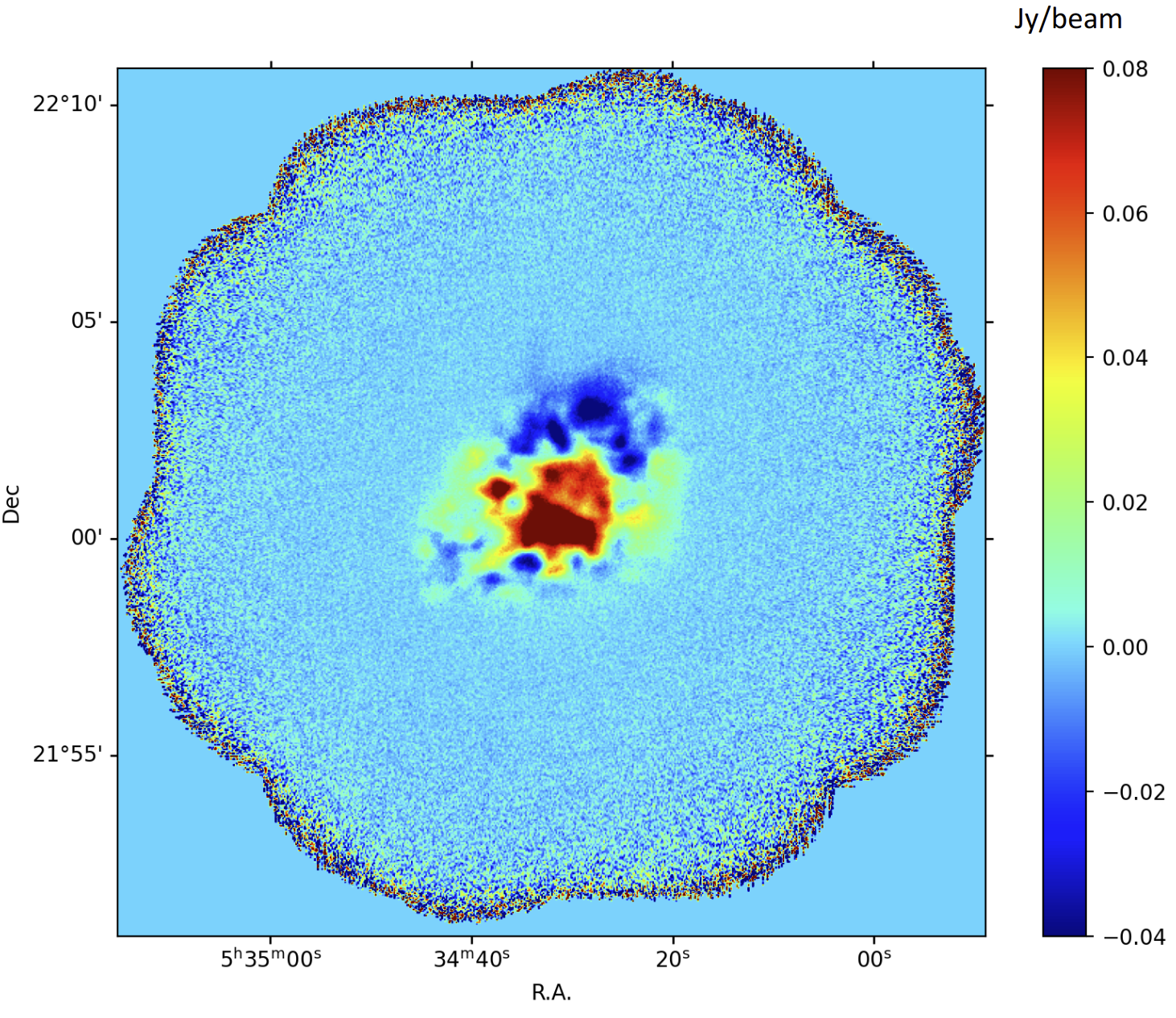}
  \includegraphics[scale=0.1]{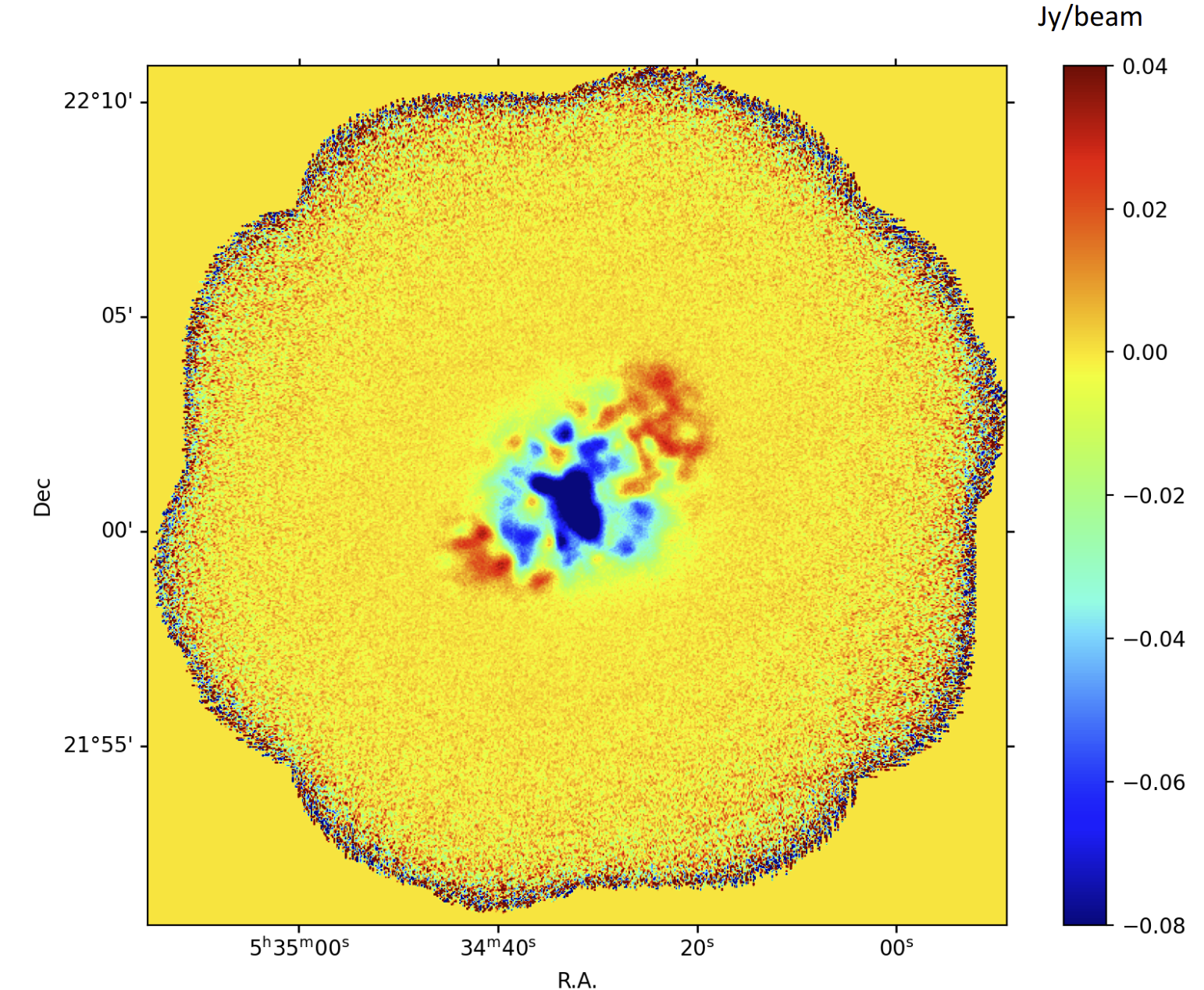}
\label{fig-1}       
\caption{From left to right: Stokes Q and U maps of the Crab nebula observed at 260 GHz during the commissioning campaign of November 2020.}
\end{center}
\end{figure*}

\begin{figure*}[h]
\begin{center}
  \includegraphics[scale=0.1]{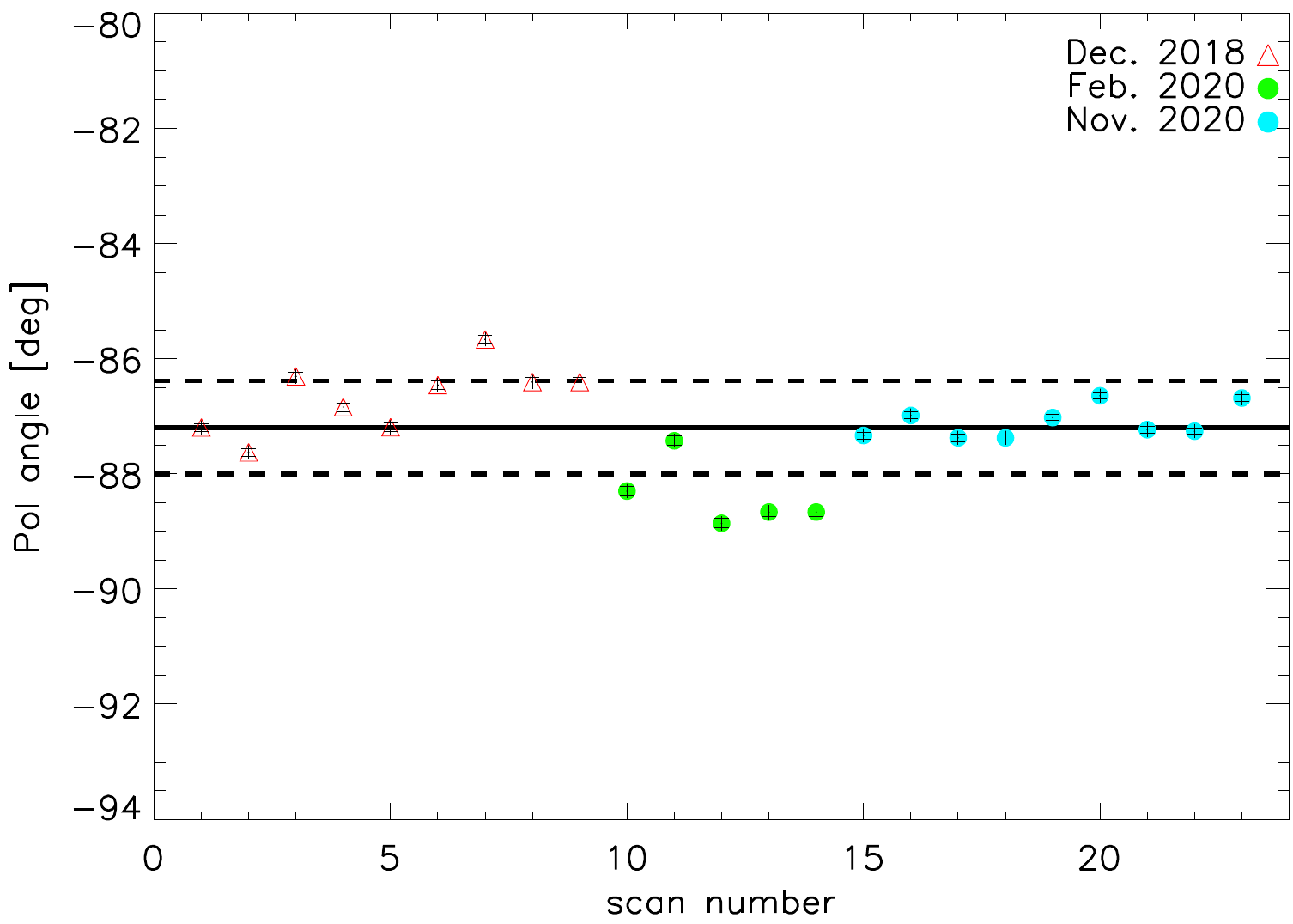}
  \includegraphics[scale=0.1]{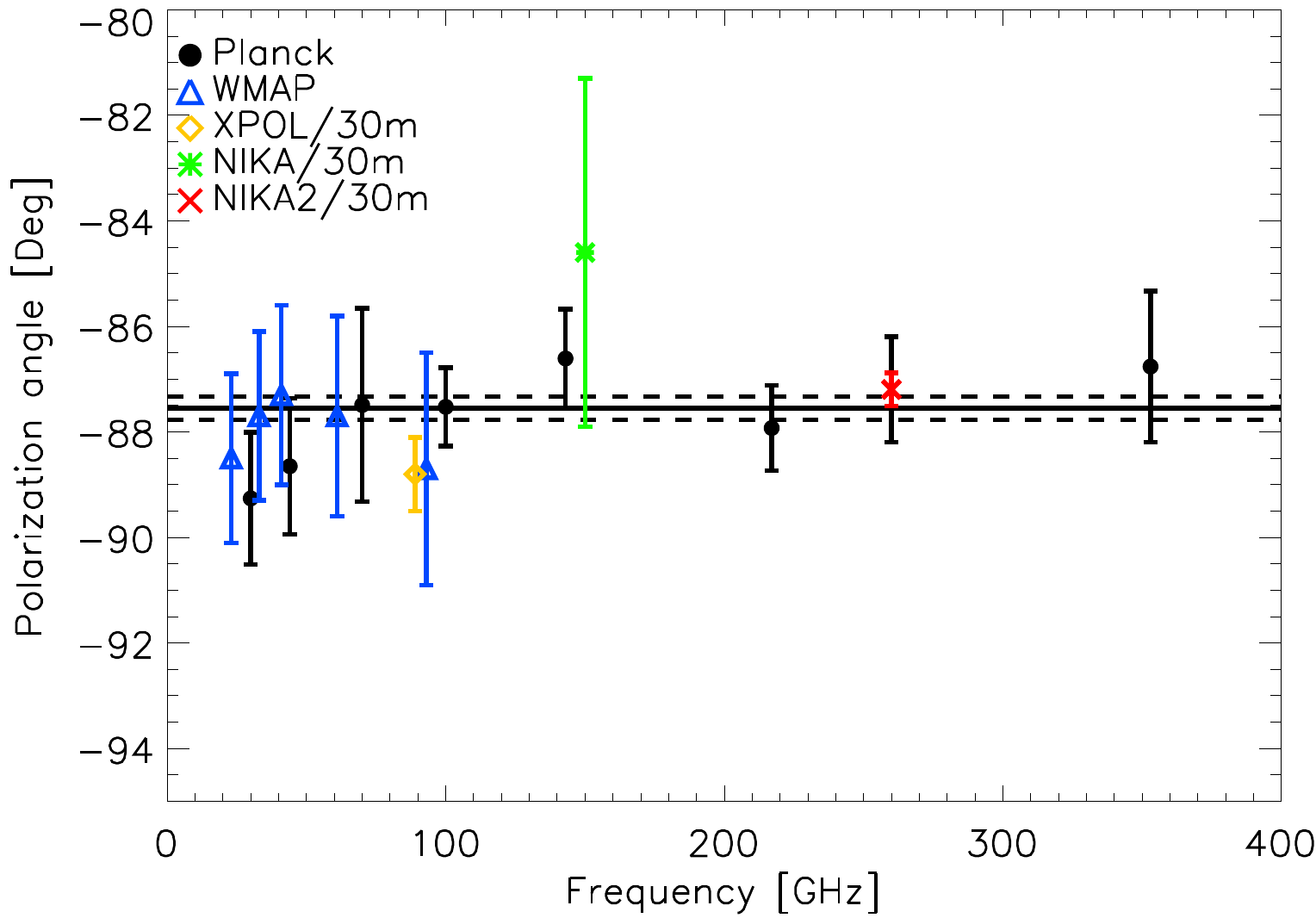}
\label{fig-2}       
\caption{Left: Variation of polarization angles measured with NIKA2-POL for the mean angle
over the Crab nebula during three commissioning campaigns. The thick black line shows the median value, the dashed lines delineate the 0.8$^{\circ}$ standard deviation over all scans, spanning $\sim$ 2 years. Right: NIKA2-POL mean polarization angle over the Crab nebula, obtained in the last campaign, vs other experiments \cite{crabnika}. The red error bar represents the uncertainty on the NIKA2 value, computed as dispersion within the observing scans of Nov. 2020, and the
black error bar shows the absolute 1$^{\circ}$ uncertainty due to the absolute angle calibration offset.}
\end{center}
\end{figure*}

%
%
%
\section{Polarized Spectral Energy Distribution}
\begin{figure}
\begin{center}
  \includegraphics[scale=0.1]{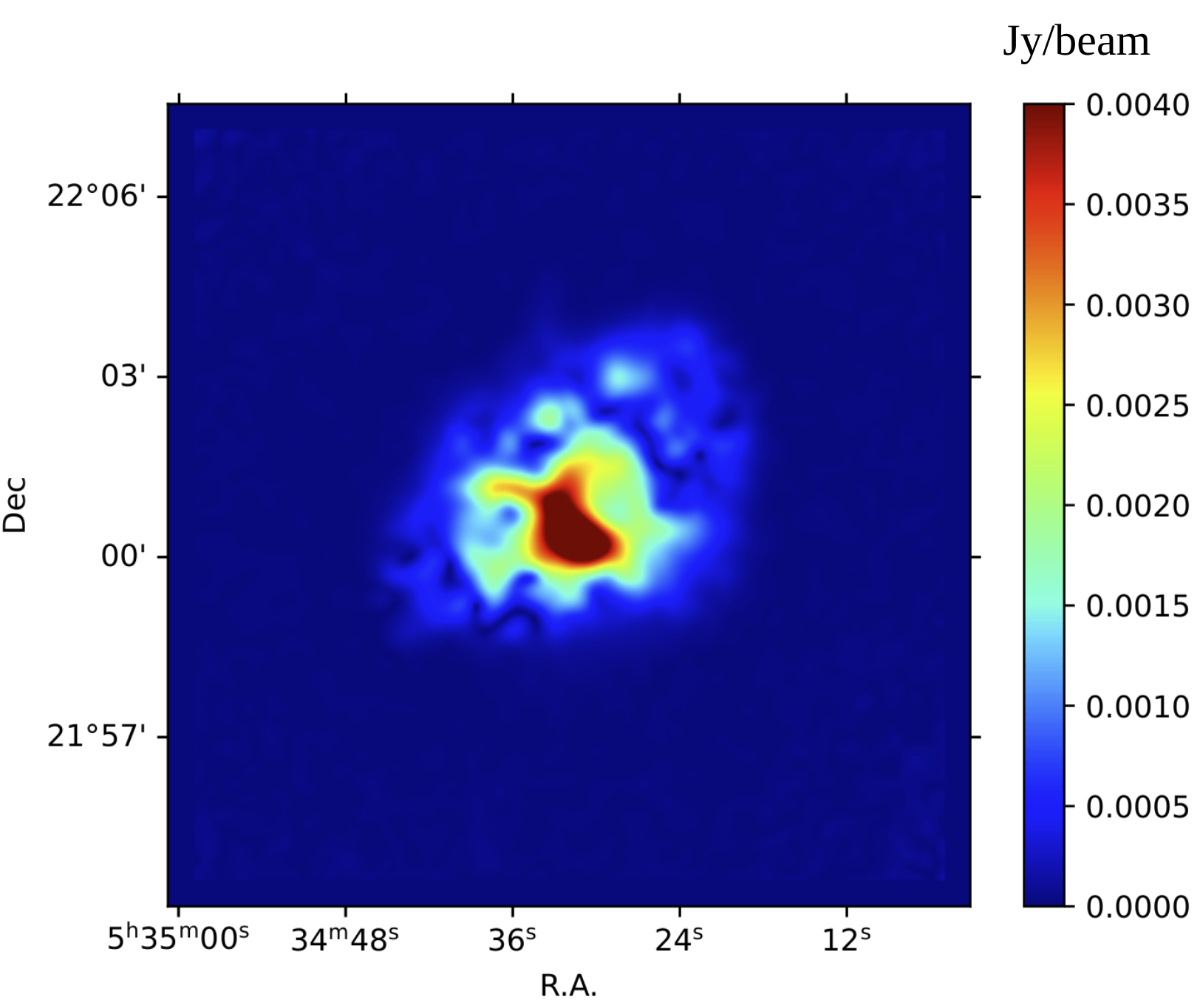}
  \includegraphics[scale=0.11]{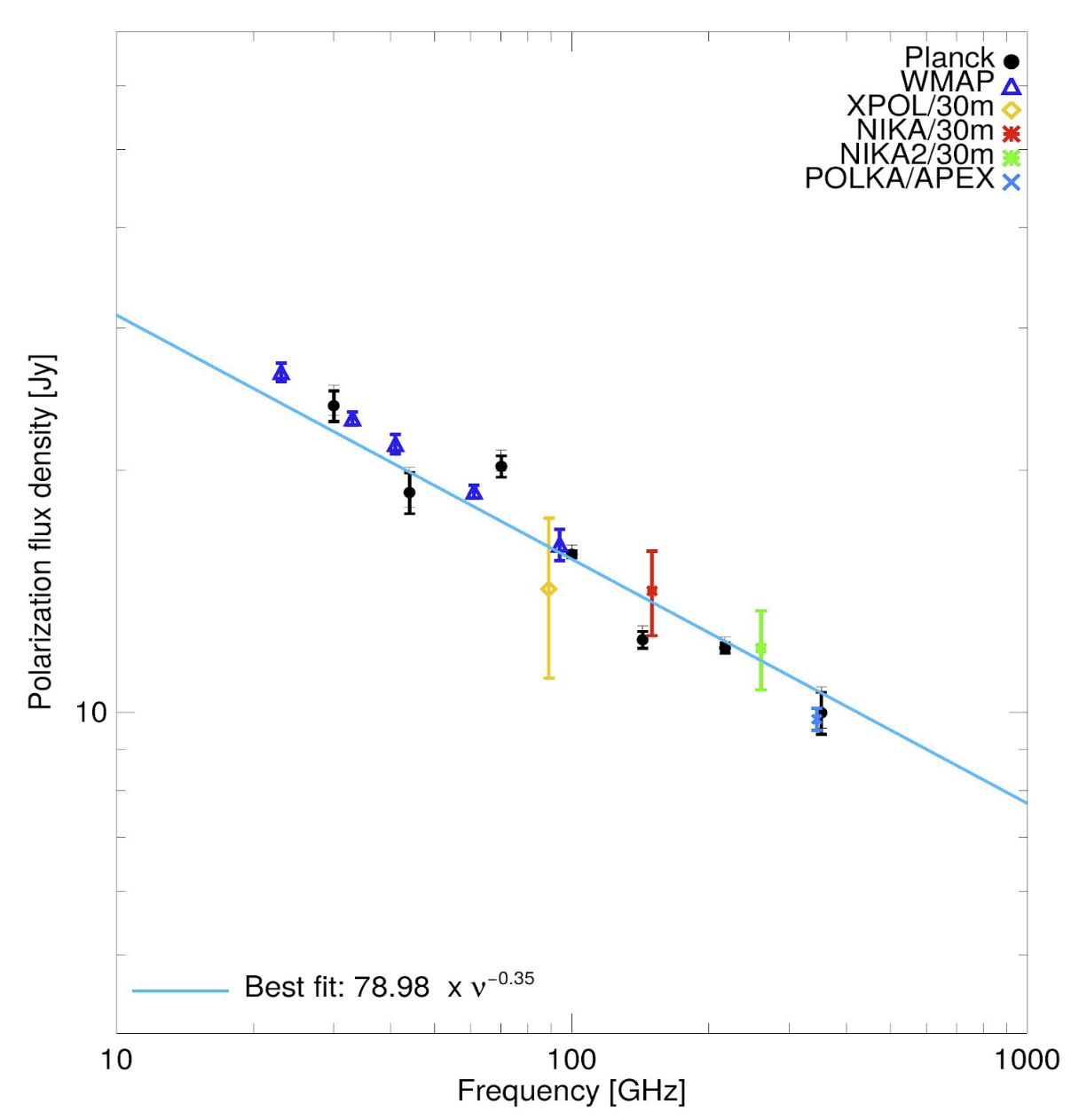}
\label{fig-3}       
\caption{Left: NIKA2 polarized intensity $P$ map. Right: spectral energy distribution obtained by previous measurements \cite{crabnika} accounting for the new value obtained from NIKA2 (green).}
\end{center}
\end{figure}

The polarized spectral energy distribution of the Crab nebula has been determined for the first time in \citep{crabnika}. Figure~\ref{fig-3} left panel shows the polarized intensity map obtained as $P = \sqrt{Q^2+U^2}$ at 260 GHz with NIKA2 and on right the spectral energy distribution as estimated by \cite{crabnika} accounting also for the new value estimated with observations from NIKA2 (green), which is P = 13.22 $\pm$ 0.15 (dev. standard) $\pm$ 1.2 (10\% calibr. error). We notice that the polarized flux measured is in very good agreement with expectations. In polarization there is no filtering of the large angular scales \citep{NIKA2-performance} and thus, the whole intensity is recovered.

\section{CMB experiment absolute calibration challenge}
\begin{figure}
\begin{center}
  \includegraphics[scale=0.1]{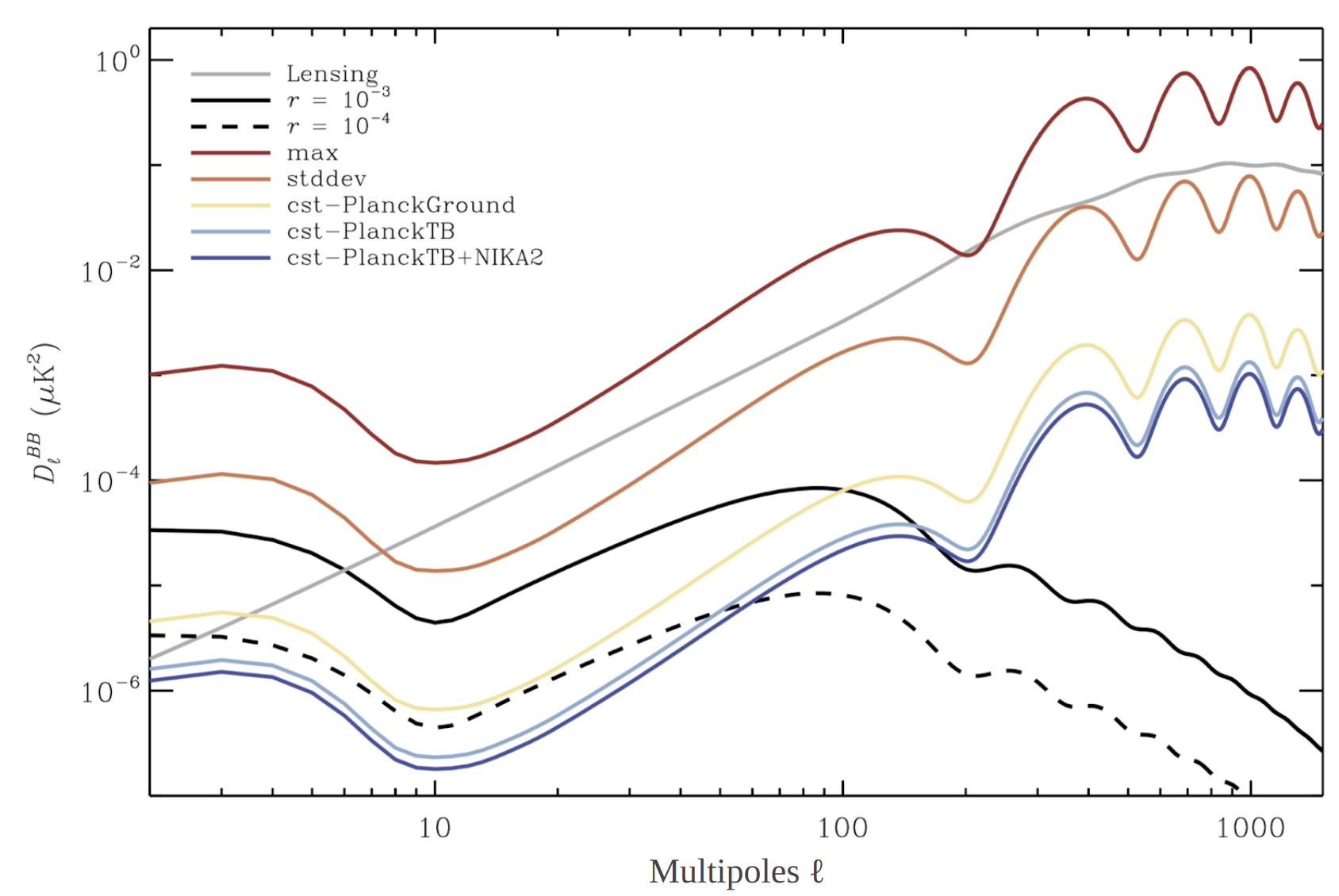}
  \includegraphics[scale=0.153]{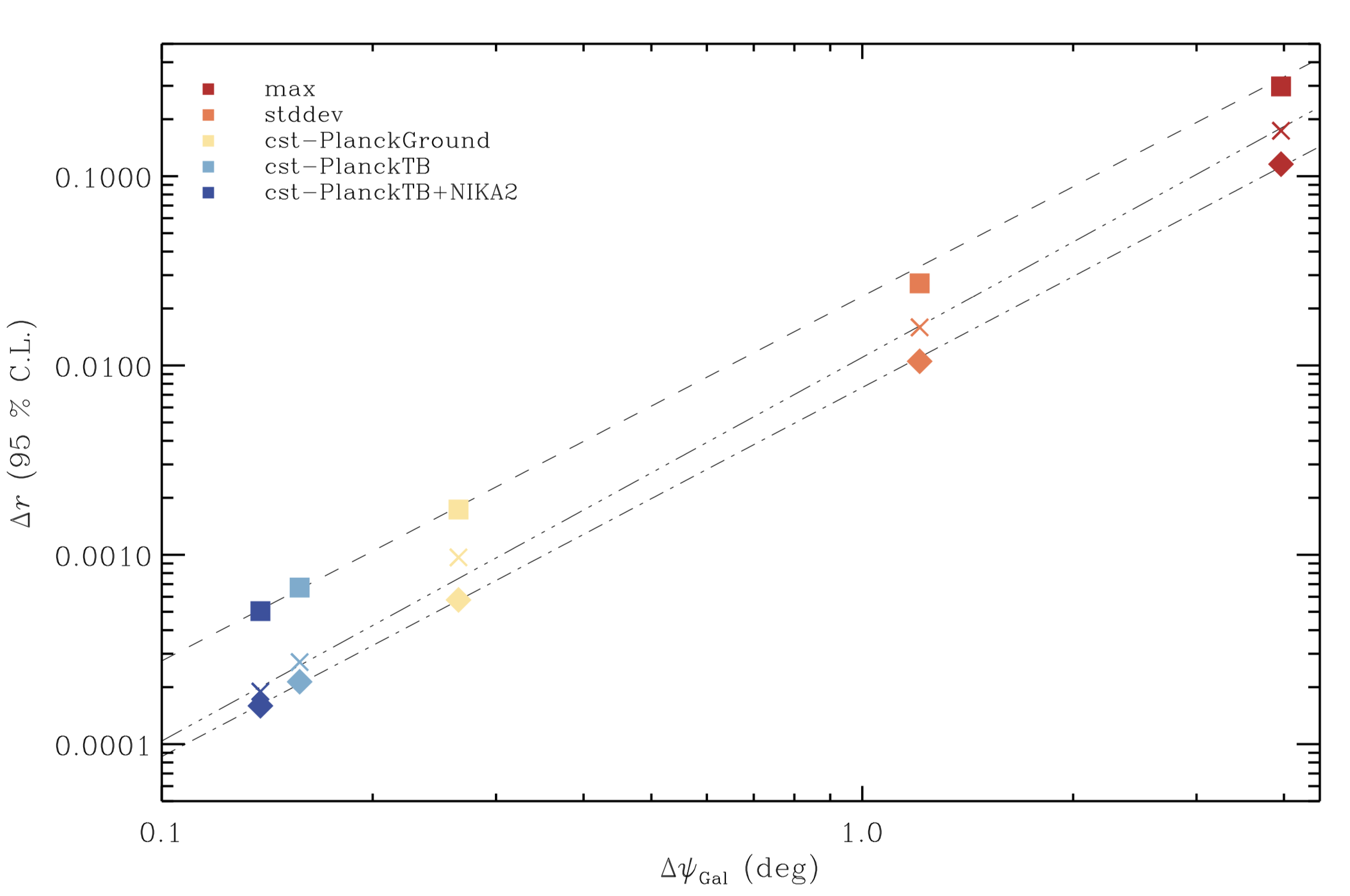}
\label{fig-4}       
\caption{Left panel: power spectrum bias from $E-B$ mixing due to the miscalibration of the absolute polarization angle. 
This bias is plotted for the different absolute calibration errors $\Delta \psi$ presented in \cite{aumont2020}. The $\Lambda$CDM best fit $D_\ell^{BB}$ primordial tensor model for $r = 10^{-3}$ and $r = 10^{-4}$ (solid and dashed black lines, respectively) and $D_\ell^{BB}$ lensing model (gray line) are also displayed. Right panel: same as left panel, but relative to the primordial tensor model for $r = 10^{-3}$.}
\end{center}
\end{figure}

The stability of the polarization angle with frequency is the key to use the Crab nebula as a sky calibrator.
The absolute calibration of the polarization angle is of a particular importance for a CMB experiment targeting the measurement of the primordial gravitational waves through the CMB polarization B-modes. Indeed, a miscalibration of an angle $\Delta\psi$ is equivalent to a rotation of the polarization plane by the same angle, mixing the true linear polarisation Q and U and consequently the polarized angular power spectra that describe the amplitude of the signal.
In the CMB and because the CMB polarization E-modes power spectrum C$_{l}^{EE}$ is greater than B-modes one C$_{l}^{BB}$, this is often referred to as an ``E to B leakage'' and reads:

\begin{equation}
  \tilde{C}_\ell^{BB} = C_\ell^{BB}\cos^2{2\Delta\psi}+ C_\ell^{EE}\sin^2{2\Delta\psi}
\Leftrightarrow\Delta C_\ell^{BB} \simeq (2\Delta\psi)^2C_\ell^{EE}
\label{eq:eb-mixing}
\end{equation}
where $\Delta C_\ell^{BB}$ represents the spurious bias component.
In order to use the Crab nebula polarization angle $\psi$ as an absolute angle calibrator for CMB experiments, we are interested in its uncertainty $\Delta\psi$. In the following we present a preliminary results on the uncertainty on the parameter $r$ as depending by the uncertainty on $\psi$ updating a previous study \cite{aumont2020} with the new value given by NIKA2 at 260 GHz. 
Notice that we make the strong assumption that we would be able to reduce the uncertainty on the polarization angle calibration offset of NIKA2 to a negligible value. In other words, we assume here that the NIKA2 polarization angle uncertainty is only to instabilities of the instrument itself.
Figure~\ref{fig-4} shows on left panel the $\Delta D_\ell^{BB} = \ell(\ell+1)/\left(2\pi\right)\Delta C_\ell^{BB}$ power spectrum bias from $E-B$ mixing due to miscalibration of the absolute polarization angle. On the right panel we show the likelihood analysis on the $r$ parameter. 
The different cases considered are taken from \cite{aumont2020} updating the two plots with the value measured at 260 GHz by NIKA2 in its best configuration taking only the data from the November 2020 commissioning campaign.

Figure~\ref{fig-4} shows that NIKA2 polarization angle stability improves the uncertainty $\Delta r$ by 30\% w.r.t previous studies \cite{aumont2020} enabling to approach the requirements for the measurement of the $r$ parameter with an uncertainty $\Delta r = 5 \times 10^{-4}$. This is promising for CMB experiments development but we need to carefully assess the NIKA2 absolute polarization angle offset.

\section{Conclusion}
In this paper we show the results obtained on Crab nebula observations during the commissioning phase of the NIKA2-Pol instrument. Comparing the three commissioning campaigns across 2018 and 2020 we have found that the data sets on the Crab and others sky calibrators, e.g. Orion OMC-1 and 3c286, are consistent with an absolute polarization angle calibration offset of -5.6 $\pm$ 1 deg. This systematic uncertainty prevents us, for instance, to use NIKA2 Crab nebula polarization angle value as reference for CMB experiments.
Though, the NIKA2 Crab nebula observations appears to be very consistent between one scan to another, especially in optimal weather conditions reducing dramatically the statistical error associated. Propagating this value to the likelihood analysis in order to estimate the total uncertainty on the detection of the $r$ parameter for the CMB B-modes detection, we see that we could reach the requirements for the next generation CMB experiments. This is a strong assumption on NIKA2 polarization performances and we are currently investigating on possible ways to reduce the uncertainties on the NIKA2 polarization angle offset.

\section*{Acknowledgements}
\input{acknowledgements}

%
%
%

\end{document}

%% file: acknowledgements.tex
\small{We would like to thank the IRAM staff for their support during the campaigns. The NIKA2 dilution cryostat has been designed and built at the Institut N\'eel. In particular, we acknowledge the crucial contribution of the Cryogenics Group, and in particular Gregory Garde, Henri Rodenas, Jean Paul Leggeri, Philippe Camus. This work has been partially funded by the Foundation Nanoscience Grenoble and the LabEx FOCUS ANR-11-LABX-0013. This work is supported by the French National Research Agency under the contracts "MKIDS", "NIKA" and ANR-15-CE31-0017 and in the framework of the "Investissements d’avenir” program (ANR-15-IDEX-02). This work has benefited from the support of the European Research Council Advanced Grant ORISTARS under the European Union's Seventh Framework Programme (Grant Agreement no. 291294). F.R. acknowledges financial supports provided by NASA through SAO Award Number SV2-82023 issued by the Chandra X-Ray Observatory Center, which is operated by the Smithsonian Astrophysical Observatory for and on behalf of NASA under contract NAS8-03060.}